\documentclass[review]{elsarticle}
\usepackage{lineno}
\usepackage{tikz}
\usepackage{tikz-qtree}
\usepackage[utf8]{inputenc}
\usepackage{pdflscape}
\usepackage{graphicx}
\usepackage{amssymb}
\usepackage{mdframed}
\usepackage{tcolorbox}
\usepackage{multirow}
\usepackage{pgf-pie}
\usepackage[breaklinks=true]{hyperref}
\usepackage{breakcites}
\usepackage[top=1.3in, bottom=1.5in, left=1.2in, right=1.2in]{geometry}

\usepackage{color, colortbl}
\definecolor{Gray}{gray}{0.9}

\journal{Computer Science Review}
\biboptions{authoryear}

\begin{document}

\begin{frontmatter}

\title{A Survey on Legal Question Answering Systems}

\author{Jorge Martinez-Gil}
\address{Software Competence Center Hagenberg \\ Softwarepark 32a, 4232  \\ Hagenberg im Muhlkreis, Austria \\ \url{jorge.martinez-gil@scch.at}}

\begin{abstract}
Many legal professionals think that the explosion of information about local, regional, national, and international legislation makes their practice more costly, time-consuming, and even error-prone. The two main reasons for this are that most legislation is usually unstructured, and the tremendous amount and pace with which laws are released causes information overload in their daily tasks. In the case of the legal domain, the research community agrees that a system allowing to generate automatic responses to legal questions could substantially impact many practical implications in daily activities. The degree of usefulness is such that even a semi-automatic solution could significantly help to reduce the workload to be faced. This is mainly because a Question Answering system could be able to automatically process a massive amount of legal resources to answer a question or doubt in seconds, which means that it could save resources in the form of effort, money, and time to many professionals in the legal sector. In this work, we quantitatively and qualitatively survey the solutions that currently exist to meet this challenge. \\ 
\end{abstract}


\end{frontmatter}

%\linenumbers

\section{Introduction}
Legal Intelligence (LI) is a broad umbrella covering a group of technologies aimed at automating tasks that formerly required human intelligence, communication, and decision-making in the legal domain. Many applications use LI technology to support services, including legal decision support, adaptive interfaces, and mining litigation data for strategic legal advice, etc. In the context of this work, we focus on a subset of the Question Answering (QA) challenge; the design of solutions for automatic Legal Question Answering (LQA). The LQA challenge is about building methods and tools that involve accurately ingesting the query information from a person as an input and automatically offering a response tailored to the person's needs as an output. 

The LQA challenge is far from being trivial since one of the legal domain's significant problems is that newly generated legislation is usually formatted unstructured. Moreover, the vast volume and speed at which legislation is made available usually lead to an information overload problem. Therefore, finding practical solutions is considered a significant research challenge, and it is a matter of intense research. In fact, in recent times, we have witnessed an explosion in the number of solutions for LQA. 

The truth is that this domain is a perfect candidate for data and knowledge scientists and practitioners to validate their systems because it meets all the requirements that make the problem attractive: there is a massive amount of information, which also grows and grows every day (e.g., the vast amount of legislation at local, regional, national and international levels that is generated every day), and this information often lacks structure (almost all public administrations publish their laws in unstructured formats and without annotation, at most with some metadata). As a result, people in the sector face severe difficulties navigating this sea of information.

Numerous solutions claim to solve the problem to overcome the above-mentioned problems (with varying degrees of success). This is mainly because these solutions are explicitly designed to automatically process many legal sources to answer a question or doubt in a matter of seconds, which means it could save resources in the form of effort, money, and time for many legal professionals. Therefore, both academia and industry have a strong interest in presenting their solutions as an excellent candidate to overcome some of the limitations in the legal field by trying to automatically provide answers to given questions, rather than presenting the professionals with a list of related items \citep{key-Kolomiyets}.

If we look at the historical evolution of this type of solution, we can realize that in the past, most techniques were based on the development or refinement of information retrieval (IR) techniques that would allow legal texts to be processed to answer questions. However, in recent times, both academia and industry are turning more to neural network-based solutions. However, both have significant advantages and disadvantages. For example, while classic IR techniques require very long processing times, which can be mitigated with high-performance hardware architectures, neural network-based techniques require a lot of training time, but once trained, they are much faster, and they usually deliver good results. However, the significant problems associated with these neural architectures are the low interpretability of the final models, the tremendous amount of resources needed for their training, and the great difficulty in transferring the learned models.

The tremendous impact that QA systems can have on a wide range of academic disciplines and business models has attracted much attention from both the research community and the legal industry. As a result, there already several surveys covering several aspects of this challenge, e.g., \citep{key-Wang,key-Kolomiyets,key-Hoffner,key-Diefenbach,key-Franco,key-Dimitrakis,key-Silva}. However, the present survey is different from the previous ones in many ways. First, it is, to the best of our knowledge, the first attempt to compile the existing literature to address the LQA problem. Moreover, we intend to do it informative and transverse, without focusing on a specific computational method or family of information sources. Finally, it is the most updated to date. We are collecting the latest advances related to the advances in neural computation and knowledge graphs.

Therefore, to articulate this survey's organization, we intend to answer the five questions that we consider key to meeting the challenge. This means that the contribution of the present work is based on an objective and depth answer to the following five main research questions:

\begin{itemize}
	\item RQ1. What are the most significant LQA research works to date?
	
	\item RQ2. How do the primary data sources on which LQA relies look like?

	\item RQ3. How do the current LQA solutions address the problem of interpretability?
	
	\item RQ4. How is the quality of LQA systems currently assessed?
	
	\item RQ5. What are the research challenges that remain open in this context?
\end{itemize}

The rest of the work is structured as follows: Section 2 addresses RQ1 by presenting the state-of-the-art LQA methods categorized by the different conceptual approaches they are based on. Section 3 addresses RQ2 by explaining the primary sources of data, information, or knowledge from which existing solutions are usually sourced. Section 4 addresses RQ3, introducing the concept of interpretability, explaining its paramount importance in the legal domain, and analyzing which solutions are the most advanced in this respect. Section 5 addresses RQ4, explaining which are the most common methods for the evaluation and quality assessment of the different LQA systems. Section 6 addresses RQ5 explaining the challenges that industry and academia will face soon if they want their solutions adopted by professionals in the sector. Finally, we note the conclusions and lessons learned that could be drawn from this work.

\section{Related Works}
In the context of this work, we assume that a question usually comes in the form of a natural language sentence, which usually starts with an interrogative word and expresses some information need of the user, although sometimes a question could also take a form of an imperative construct and starts with a verb. The answer to a question, which may or may not be known in advance, would be a word, number, date, sentence, or paragraph that completely satisfies the information need expressed by the question. The standard process of automatically answering a question involves three fundamental phases: Query Understanding (to extract the core elements that make up the question), Sources Retrieval (to identify a few documents that may contain the answer from a large source pool), and Answer Extraction (to find a short phrase or sentence that answers the user’s question directly). However, not all methods implement the different phases similarly, as we will see throughout the section.

The problem of automatically match an answer to a question is widely assumed to be an important research challenge that tries to address one aspect of Artificial Intelligence (AI) that will allow computers to perform routine and tedious tasks. Nevertheless, automatically answering questions is genuinely complex and encompasses many challenges and nuances that make it unfeasible to be covered by just an application. This means that most of the associated challenges could have an impact on both industry and academia. This is because having information systems that can correctly answer questions asked by a person opens a wide range of possibilities to impact, from the most basic research to the most advanced business models.

For this reason, there are numerous solutions to the problem of QA. While in the past, most techniques were based on the development or refinement of natural language techniques that would allow legal texts to be processed to answer questions, in recent times, both academia and industry are turning more to neural network-based solutions inspired in the seminal work of Mikolov et al. \citep{key-Mikolov}. For example, BERT \citep{key-BERT} and ELMo \citep{key-elmo} or USE \citep{key-Cer}. These approaches have tremendous advantages and disadvantages. For example, while traditional techniques require properly tuning up different workflow pipelines, neural network-based techniques can be trained once and use many times later. The problem is the lack of interpretability, i.e., the limited capacity a person might have to understand why the methods work so well and the number of resources required for training and adaptation to new cases.

In this survey, we focus strictly on existing LQA work, i.e., questions and answers restricted to the world of law. Table 1 shows a summary of the features that make them different from other question answering systems. First of all, in LQA, we work in a closed environment. Moreover, the questions can only be related to the legal domain. In addition, all types of questions are allowed, whether they are definition, comparison, confirmation, etc. The analysis of the texts, both questions and sources to extract the answers, is not limited. Nor are the types of sources that can be used to find the answer limited. Last but not least, in the context of the current survey, we do not include dialogue systems but only question answering solutions.

\begin{table}[]
\centering
\caption{Kind of QA systems covered within this survey}
\begin{tabular}{ lll }
\hline
\multicolumn{3}{ c }{Features of Legal Question Answering} \\
\hline \hline
\rowcolor{Gray}{Environment} & [Open] & \textbf{Closed} \\
\rowcolor{Gray}\multirow{-1}{*} & [Closed] &  \\ \hline 
\multirow{3}{*}{Domain} & [General-purpose] & \textbf{Legal} \\
 & [Mathematical] &  \\
 & [Legal] &  \\
 & [Other] &  \\ \hline 
\rowcolor{Gray}{Kind of Questions} & [Factoid] & \textbf{All}\\
\rowcolor{Gray} & [Definition] &  \\
\rowcolor{Gray} & [Confirmation] &  \\
\rowcolor{Gray} & [Causal] &  \\
\rowcolor{Gray}\multirow{-4}{*} & [Comparison] & \\ \hline
\multirow{3}{*}{Kind of Analysis} & [Morphological] & \textbf{All} \\
 & [Syntactical] &  \\
 & [Semantic] &  \\
 & [Pragmatic]&  \\ \hline \hline 
\rowcolor{Gray}{Sources} & [Structured] & \textbf{All} \\
\rowcolor{Gray} & [Semi-structured] &  \\
\rowcolor{Gray}\multirow{-2}{*} & [Unstructured] & \\ \hline
\multirow{3}{*}{Kind of dialogue} & [Question Answering] & \textbf{Question Answering} \\
 & [Dialog] & \\
 & [Spoken Dialog] & \\ \hline
\hline
\end{tabular}
\end{table}

Several different classifications schemes can be potentially established for categorizing the main families of systems built in this domain. QA is a form of information retrieval, and this means that solutions are defined by how they represent the input information and the retrievable entity as well as by the ranking function. To properly articulate this section, we have focused on the different computational approaches for building LQA solutions. Techniques of this kind have been widely used in various forms of research on Yes/No Answers, Multiple Choice Question Answering, Classical Solutions from the Information Retrieval field, Ontology-based solutions, Neural solutions, and Other approaches. 

\begin{itemize}
	\item Yes/No Answers
	\item Legal Multiple Choice Question Answering
	\item Classical IR-based Solution
	\item Big Data solutions
	\item Ontology-based solutions
	\item Neural solutions
	\item Other approaches, e.g., Knowledge Graphs
\end{itemize}

Each of the different branches of research has its peculiarities. Thus, for example, the first types based on Yes/No Answers and Multiple Choice Question Answering assume a simplified version of the problem where it is unnecessary to worry about generating candidate answers, only about confirming or not the veracity of a sentence or constructing a ranking of answers. Classical Information Retrieval techniques, in principle, should be one the most numerous families since they are one of the natural ways of tackling the problem. On the other hand, Big Data techniques, even if they do not represent an apparent scientific breakthrough, do represent a considerable advance in technical characteristics, where even unfeasible but straightforward techniques on a regular computer (think, for example of the calculation of regular expressions) can have a good performance. Finally, the novel techniques based on neural architectures with the best-expected results and other approaches come to fill some of the gaps related to neural networks, such as, for example, the lack of interpretability. We then explain what search strategy we followed to capture the primary sources of the literature, we review the different approaches, and we conclude the section with a summary.

\subsection{Data Sources and Search Strategy}
To choose the primary sources of this survey, we have built a search string so that it includes two major search terms:  \textit{'Method'} and \textit{'Field'}. The first major search term represents the employed methodology, namely, 'question answering', whereas the second major search term illustrates the fields in which the method should have been utilized, i.e., 'legal'. This term includes all sorts of technologies and synonyms of legal in which that application should occur. Terms like 'legal domain', 'legal sector', 'legal industry', 'legal area', 'legal documents', 'legal material', 'legislation', and 'regulatory'. In addition, through snowballing, we have come up with articles that contain terms such as 'due diligence'. Queries were run on Google Scholar\footnote{https://scholar.google.com/}, DBLP\footnote{https://dblp.org/}, Microsoft Academic\footnote{https://academic.microsoft.com/home}, and Semantic Scholar\footnote{https://www.semanticscholar.org/}.

Moreover, we have had to use some criteria to decide whether a particular work should be considered. Table 2 shows the criteria we have followed in deciding whether a paper should be considered for inclusion or exclusion in this survey. This table follows the recommendation stated by \citep{key-Soares} for the study of QA systems.

\begin{table}[]
\centering
\caption{Inclusion and exclusion criteria for work selection}
\begin{tabular}{ll}
\hline 
\textbf{Inclusion} & \textbf{Exclusion}\\
\hline \hline
\rowcolor{Gray}
Written in English    &    Written in other languages rather than English     \\
Published in journal, conference or workshop  &  Preprints when the paper is available      \\
\rowcolor{Gray}
Described an algorithm, technique or system  & Thesis, editorials, prefaces, or summaries  \\
Uses, at least, one LQA dataset  &   QA method but no application in LQA \\
\rowcolor{Gray}
Published from January 1990	&  Published by a predatory publisher \\
Published until June 2021	&  Patents or patent applications \\
\hline     
\end{tabular}
\end{table}

As a result of the screening process, we have obtained 65 primary sources analyzed and categorized below within their corresponding family of solutions.

\subsection{Yes/No Answers}
This approach is the simplest of all. Given a question and a known answer, one tries to determine whether the associated answer is true or false. In this way, the systems do not need to generate candidate answers. Nevertheless, answering yes/no questions in the legal area is very different from other domains. One of its inherent features is that legal statements need a thorough examination of predicate-argument structures and semantic abstraction in these statements. The methods from this family usually check if there is a high degree of co-occurrence between questions and answers in corpora of a legal nature. Some variants might result in searching for some kinds of regular expressions.

Works based on this approach, such as \citep{key-Kim} have developed different approaches to answer yes/no questions relevant to civil laws in legal bar exams. A bar examination is intended to determine whether a candidate is qualified to practice law in a given jurisdiction. There is a recurring concern in the literature on whether it is possible to pass this type of test without human supervision. The same author \citep{key-Kim2} has also implemented some unsupervised baseline models (TF-IDF and Latent Dirichlet Allocation (LDA)) and a supervised model, Ranking SVM, for facing the challenges. The model features are a set of words and scores of an article based on the corresponding baseline models. In addition, a final enhancement includes the use of paraphrasing \citep{key-Kim4}. Other authors \citep{key-Taniguchi,key-Taniguchi2} has addressed the problem by case-role analysis and Framenet\footnote{https://framenet.icsi.berkeley.edu/fndrupal/}, respectively. Finally, Kano et al. has explored, for the first time, linguistic structures \citep{key-Kano}. Table 3 summarizes all these works.

\begin{table}[]
\centering
\caption{Summary of approaches based on Yes/No}
\begin{tabular}{ll}
\hline
\textbf{Work} & \textbf{Approach }\\
\hline \hline
\rowcolor{Gray}
\citep{key-Kim}     & Antonym detection and Semantic Analysis    \\
\citep{key-Kim2}   & TF-IDF, LDA and SVM  \\
\rowcolor{Gray}
\citep{key-Kim4}   & Paraphrasing detection \\
\citep{key-Taniguchi}   & Case-role analysis  \\  
\rowcolor{Gray}
\citep{key-Kano}   & Using linguistic structures  \\  
\citep{key-Taniguchi2}   & Using Framenet  \\ 
\hline     
\end{tabular}
\end{table}

This branch of research was one of the first to yield results. Its strength is based on a superficial understanding of the problem with solutions that tend to work well. The major problem of this model is that questions in human language express a well-defined information requirement on the one hand, but they also convey more information than an essential list of relevant keywords since they represent syntactic and semantic links between those keywords on the other. Therefore, solutions tend to work well only to a limited extent, i.e., when the questions are simple.

\begin{tcolorbox}
\noindent \textbf{Pros:} High interpretability. Methods can be used in many other domains for fact-checking \\
\noindent \textbf{Cons:} The operating model is trivial and leads to poor results at the present time
\end{tcolorbox}

\subsection{Legal Multiple Choice Question Answering}
Legal Multiple Choice Question Answering (LMQA) consists of correctly answer questions in a scenario where the possible answers are already given beforehand. In this model, there are two clear facilitators. It is unnecessary to generate candidate answers since this family of methods assumes that candidate answers are a different problem and should be studied apart. Furthermore, it is known that the correct answer is among the given ones (even in situations where \textit{None of above} answers are allowed). Therefore, the goal is to learn a scoring function $S(F, z)$ with a normalization parameter $z$ (whereby $z$ or the normalization factor is usually used so that the values associated with each answer are in the same numerical range) such that the score of the correct choice is higher than the score of the other hypotheses and their corresponding probabilities.

Some works have used the classical Latent Semantic Analysis to obtain good results in this context \citep{key-Deerwester}. In general, all techniques based on co-occurrence are very appropriate because it is usually sufficient to check the number of times that the question and each of the answers appear together in some textual corpora of a legal nature \citep{key-Li}. Of course, one improvement is to use semantic similarity detection techniques to detect different formulations of the same information \citep{key-sts}. In addition, crowdsourcing techniques have also been very successful \citep{key-Aydin}. There are variations of the above that attempt to compute co-occurrence in a more sophisticated way so that text windows and regular expressions are taken into account \citep{key-martinez-dexa}. More recently, \citep{key-Chitta} has proposed an improvement through multi-class classifiers to identify the answer to the question. The key idea is to handle imbalanced data by generating synthetic instances of the minority answer categories. Table 4 chronologically lists the works that we have mentioned above.
 
\begin{table}[]
\centering
\caption{Summary of approaches based on Multiple Choice}
\begin{tabular}{ll}
\hline 
\textbf{Work} & \textbf{Approach }\\
\hline \hline
\rowcolor{Gray}
\citep{key-Deerwester}    &    Latent Semantic Analysis      \\
\citep{key-Li}    &    Co-occurrence      \\
\rowcolor{Gray}
\citep{key-sts}    &    Synonym detection     \\
\citep{key-Aydin} &     Crowdsourcing      \\
\rowcolor{Gray}
\citep{key-martinez-dexa}   & Reinforced Co-occurrence  \\
\citep{key-Chitta}    &    Synthetic minority oversampling      \\  
\hline     
\end{tabular}
\end{table}

One of the fundamental keys for LMCQA to work well is the choice of the legal corpus to be worked on, both in terms of extension and quality since most methods require the calculation of semantic similarity, textual co-occurrences, etc.

\begin{tcolorbox}
\noindent \textbf{Pros:} High speed. High interpretability. Many existing methods for calculating rankings \\
\noindent \textbf{Cons:} Knowing all answers beforehand is not a realistic situation in the real world
\end{tcolorbox}

\subsection{Classical IR-based Solutions}
In this context of LQA, classical IR-based solutions usually represent words in the form of discrete and atomic units. This family of methods assumes that it is vital that the solution appropriately identify either the exact answer or source and specific paragraph containing the relevant information. The lawyer then should read and interpret the excerpt for the client. For example, the first approach (and the simplest) could query the number of Google results for a specific question and a given answer together. However, this solution has brought several problems like the lack of context for the formulated question. 

To overcome the problem of the lack of context, word processing models such as LSA \citep{key-Deerwester} and term frequency-inverse document frequency (TF-IDF) partially solve these ambiguities by using terms that appear in a similar context based on their vector representation. Then they group the semantic space into the same semantic cluster. Within this family of methods, one of the best-known QA systems is IBM Watson \citep{key-Ferrucci}, which is very popular for its victory in the televised show \textit{Jeopardy} \citep{key-Jeopardy}. Although in recent times, IBM Watson has become a generic umbrella that includes other business analytics capabilities.

If we restrict ourselves to LQA, one of the most popular initiatives is ResPubliQA. This evaluation task was proposed at the Cross-Language Evaluation Forum (CLEF) \citep{key-Penas,key-Penas2} a consists of given a pool of independent questions in natural language about European legislation, proposed systems should return a passage (but not an exact response) that answers each question. This initiative was a remarkable success. In fact, thanks to ResPubliQA, this family of methods has been the subject of intensive research during the last decade. 

In relation to the works of this family, it is possible to identify, Brueninghaus and Ashley that have addressed the problem for the first time with a focus on information extraction methods \citep{key-Bruninghaus}. Quaresma et al. focused on Portuguese legislation using a classical IR pipeline \citep{key-Quaresma}. Some years after, \citep{key-maxwell} went a step further beyond by using conceptual and contextual search. Another strategy was proposed by \citep{key-Monroy} that returns several relevant passages extracted from different legal sources. The relevant passages allows generating answers to questions formulated in plain language using graph algorithms. A qualitative improvement was presented \citep{key-Monroy2} to lemmatizing and using manual and automatic thesauri for improving question-based document retrieval. For the construction of the graph, the authors followed the approach of representing all the articles as a graph structure. In parallel, \citep{key-Tran} proposed mining reference information. Last but not least, \citep{key-Rodrigo2} proposed a new IR-based system with the lessons learned from ResPubliQA.

More recently, \citep{key-Carvalho} have introduced a new approach for lexical to discourse-level corpus modeling, and \citep{key-Bach} proposes a novel solution based on Conditional Random Fields (CRFs), which are methods that resorts on statistical modeling in order to segment and label sequence data. Furthermore, \citep{key-Kuppevelt} decided to explore how to perform network analysis and visualization as a proper way to deal with Dutch case law. Meanwhile, \citep{key-Delfino} has brought for the first time the Portuguese version of the thesaurus OpenWordNet\footnote{https://github.com/own-pt/openWordnet-PT} to the table.

The last works in this direction are those from \citep{key-Hoshino} by using predicate-argument structure, \citep{key-McElvain,key-McElvain2} over plain text and focusing on non-factoid kind of questions. Non-factoid questions are those whose answer is not directly accessible in the target document, which demands some inference and perhaps extra processing to obtain an answer. Wehnert et al. have tried a new application of the popular BM25 approach \citep{key-Wehnert}, and the latest to date, \citep{key-Verma} has had a focus on relevant subsection retrieval to answer the questions of legal nature. The story so far can be summarized in Table 5.

\begin{table}[]
\centering
\caption{Summary of approaches based on classical and novel approaches for Information Retrieval}
\begin{tabular}{ll}
\hline
\textbf{Work} & \textbf{Approach }\\
\hline \hline
\rowcolor{Gray}
\citep{key-Bruninghaus}     &     Information extraction methods    \\
\citep{key-Quaresma}     &     Classical IR pipeline      \\
\rowcolor{Gray}
\citep{key-maxwell}   & Conceptual and contextual search  \\  
\citep{key-Monroy}     &   Exploitation of Graph-based algorithms      \\
\rowcolor{Gray}
\citep{key-Monroy2}   & Exploitation of Graph-based algorithms    \\ 
\citep{key-Tran}   & Mining reference information  \\
\rowcolor{Gray} 
\citep{key-Rodrigo2}   & Lessons learned from ResPubliQA  \\  
\citep{key-Carvalho}   & Lexical to discourse level corpus modeling \\
\rowcolor{Gray}
\citep{key-Kim6}   & Cascaded textual entailment \\
\citep{key-Bach}   & Conditional Random Fields\\
\rowcolor{Gray}
\citep{key-Kuppevelt}   & Network Analysis\\
\citep{key-Delfino}   & Exploitation of word senses and relations  \\ 
\rowcolor{Gray}
\citep{key-Hoshino}   & Predicate Argument Structure  \\ 
\citep{key-McElvain}   & Non-Factoid questions \\ 
\rowcolor{Gray}
\citep{key-McElvain2}   & Non-Factoid questions \\   
\citep{key-Wehnert}   & BM25 and Elasticsearch \\
\rowcolor{Gray}
\citep{key-Verma}   & Relevant subsection retrieval \\
\hline     
\end{tabular}
\end{table}

As can be seen, this family is extensive. Moreover, in its bosom, many and very diverse proposals have been born to face the problem. Before the latest advances in Deep Learning, it was the dominant family, and its solutions were considered good until those times. However, in recent times very tough competitors have emerged.

\begin{tcolorbox}
\noindent \textbf{Pros:} High performance, High interpretability  \\
\noindent \textbf{Cons:} In recent times, accuracy has been overtaken by neural solutions
\end{tcolorbox}

\subsection{Big Data solutions}
In recent times, the arrival of novel Big Data approaches has come up with many advantages and challenges for the legal sector. The possibilities of Big Data like the effective and efficient processing of massive amounts of data have introduced many opportunities and challenges. Nevertheless, some of the limitations that have traditionally burdened this area of knowledge remain. Perhaps the most illustrative of these limitations is the inability of the systems to understand the question instead of merely using its huge computational potential to search for statistical patterns in corpora of a legal nature. 

For this reason, the existing approaches are focused on developing good ideas that allow to answer the questions asked and obviate the performance details since these can be solved with these new forms of high-performance computing. Some outstanding works in this direction are proposed by Bennet et al. with a focus on the scalability of the solution by design \citep{key-Bennett}, or Mimouni et al. for handling the problem of complex queries by working with approximate answers and richer semantics \citep{key-Mimouni}. Last but not least, there is also a novel approach based on the concept of mutual information exchange \citep{key-martinez-mutual}, which, when applied to large volumes of data, have demonstrated better performance than classic co-occurrence methods. Table 6 lists chronologically these works.

\begin{table}[]
\centering
\caption{Summary of approaches based on Big Data}
\begin{tabular}{ll}
\hline
\textbf{Work} & \textbf{Approach }\\
\hline \hline
\rowcolor{Gray}
\citep{key-Bennett}     &     Scalable architecture by design      \\
\citep{key-Mimouni}   & Approximate Answers and Rich Semantics  \\ 
\rowcolor{Gray}
\citep{key-martinez-mutual}   & Mutual Information Exchange \\  
\hline     
\end{tabular}
\end{table}

On the other hand, these applications entail processing and handling large amounts of data quickly, and they facilitate the automation of specific low-level computing operations. So in this way, most technical limitations disappear, and therefore, methods that were previously considered too expensive in terms of resource consumption in the form of time are no longer a problem. For example, the execution of queries based on the concept of regular expression that is very expensive in terms of computing time. However, the scientific difficulties to model appropriate inter-dependencies between the concepts from the questions remain, so the results are not optimal yet.

\begin{tcolorbox}
\noindent \textbf{Pros:} Very good performance, no limitations of a technical nature \\
\noindent \textbf{Cons:} Inability of the systems to properly understand the meaning of a complex question 
\end{tcolorbox}

\subsection{Ontology-based solutions}
The capacity to construct a semantic representation of the data paired with the accompanying domain knowledge is one of the significant benefits of adopting domain ontologies. Ontologies can also be used to establish connections between various sorts of semantic knowledge. As a result, ontologies can be utilized to develop various data-search strategies. The legal domain has not remained oblivious to this line of research.

Examples of works that deal with the first approach are Lame et al. proposed for the first time the integration of NLP and ontologies to solve LQA problems \citep{key-Lame}. Moreover, Xu et al. proposed a system for relation extraction and textual evidence\citep{key-Xu}. In addition, Fawei et al. \citep{key-Fawei}, in the first instance, proposed a criminal law and procedure ontology. Meanwhile, in the second instance, proposed a way to semi-automatically construct ontologies for the legal section \citep{key-Fawei2}. Another approach has consisted of Semantic Role Labeling \citep{key-Veena}, and last but not least, Kourtin et al. have opted for an RDF-SPARQL tailored design \citep{key-Kourtin}. Table 7 shows us a summary of the works mentioned above.

\begin{table}[]
\centering
\caption{Summary of approaches based on ontologies}
\begin{tabular}{ll}
\hline
\textbf{Work} & \textbf{Approach }\\
\hline \hline
\rowcolor{Gray}
\citep{key-Lame}    &     Combination of NLP and ontologies      \\
\citep{key-Xu}   & Relation extraction and textual evidence  \\
\rowcolor{Gray}
\citep{key-Fawei}   & Using a criminal law and procedure ontology  \\ 
\citep{key-Fawei2}   & Using semi-automated ontology construction \\
\rowcolor{Gray}
\citep{key-Veena}   &  Semantic Role Labeling \\
\citep{key-Kourtin}   & RDF-SPARQL tailored design \\
\hline     
\end{tabular}
\end{table}

Solutions based on ontological foundations have always promised to provide the correct answer to a question. However, in practice, things are not rather complicated. There are very few ontologies, and those that exist are hardly reusable. So it is costly and error-prone to develop legal ontological models from scratch. Not to mention the learning curve that dealing with description logics present, and the difficulty of performing large-scale reasoning in reasonable time.

\begin{tcolorbox}
\noindent \textbf{Pros:} Logical reasoning provides accurate answers. High interpretability \\
\noindent \textbf{Cons:} Models very expensive to build, very difficult to reuse, slow reasoning times
\end{tcolorbox}

\subsection{Neural solutions}
Solutions based on architectures inspired by neural models have adequately and consistently solved most benchmarks in the QA field. These results are outstanding and have made most of the community decide to use their resources to explore this direction. In recent times, several works have been presented. There are two significant lines of research: A branch that tries to develop solutions that exploit structured resources that have led to a family that we will see later. Furthermore, another branch that tries to develop methods capable of exploiting unstructured information. 

Exemplary works that operate with unstructured text using the most modern machine learning techniques are, for example, works such as \citep{key-Kim3,key-Do} used for the first time a convolutional neural network to LQA. A few years after, the same authors proposed an improvement using deep siamese networks \citep{key-Kim5}. Other authors decide to explore new kinds of systems using LSTM \citep{key-Adebayo}, the concept of neural attention \citep{key-Morimoto}. The concept of neural attention consists of freeing the encoder-decoders architecture from a fixed-length internal representation. Furthermore, the exploitation of a Multi-task Convolutional Neural Network \citep{key-Xiao} has also been proposed.

Some recent works such as \citep{key-Cer,key-BERT,key-elmo} have made some of the most critical advances in question answering in recent times. These approaches have consistently achieved top results in most academic competitions with a clear general-purpose orientation. Moreover, its performance in the legal sector is more than remarkable.

Futhermore, \citep{key-Collarana} introduced for the first time a new technique based on Match-LSTM and a Pointer Layer, whereas \citep{key-Nicula} proposed to use Deep Learning over candidate contexts. \citep{key-Liu} built a system using Deep Neural Networks. In parallel, Ravichander et al. proposed a new strategy based on the combination of computational and legal perspectives \citep{key-Ravichander}. And last, but not least, Zhong et al. proposed two novel alternatives using Reinforcement Learning and Deep Learning respectively \citep{key-Zhong,key-Zhong2}.

Among the latest published works in this family of methods, we can highlight the following: The Neural Attentive Text Representation from \citep{key-Kien}, the introduction of Few-shot Learning in the legal domain \citep{key-Wu}. Novel techniques for language modeling \citep{key-Huang}. Deep learning restricted to the world of building \citep{key-Zhong3}. And diverse applications of the successful BERT to implement LQA solutions \citep{key-Holzenberger, key-Zhang}. Table 8 shows a chronological summary of the aforementioned solutions.

\begin{table}[]
\centering
\caption{Summary of LQA approaches based on neural networks}
\begin{tabular}{ll}
\hline
\textbf{Work} & \textbf{Approach }\\
\hline \hline
\rowcolor{Gray}
\citep{key-Kim3}     &     Convolutional Neural Network     \\
\citep{key-Adebayo}     &     Long-Short Term Memory     \\
\rowcolor{Gray}
\citep{key-Do}     &     Convolutional Neural Network     \\
\citep{key-Kim5}   & Deep Siamese Networks \\
\rowcolor{Gray}
\citep{key-Morimoto}   & Neural Attention  \\
\citep{key-Xiao}   & Multi-task Convolutional Neural Network \\ 
\rowcolor{Gray} 
\citep{key-Cer} 	& Universal Sentence Encoder \\
\citep{key-BERT}   & Bidirectional Encoder Representations from Transformers \\ 
\rowcolor{Gray} 
\citep{key-elmo}   & Embeddings from Language Model  \\ 
\citep{key-Collarana}   & Match-LSTM + Pointer Layer \\
\rowcolor{Gray}
\citep{key-Nicula}   & Deep Neural Networks over candidate contexts  \\ 
\citep{key-Liu}   & Deep Neural Networks \\
\rowcolor{Gray}
\citep{key-Ravichander}   & Combination of computational and legal perspectives \\
\citep{key-Zhong}   & Reinforcement Learning \\
\rowcolor{Gray}
\citep{key-Zhong2}   & Deep Neural Networks \\
\citep{key-Kien}   & Neural Attentive Text Representation \\ 
\rowcolor{Gray} 
\citep{key-Wu}   & Few-shot Learning \\
\citep{key-Holzenberger}   & Based on BERT \\
\rowcolor{Gray}
\citep{key-Huang} & Language Modeling \\ 
\citep{key-Zhong3}   & Deep Neural Networks over building regulation \\
\rowcolor{Gray}
\citep{key-Zhang}   & Based on BERT \\
\hline     
\end{tabular}
\end{table}

The advances in the numeric representation of words has opened a complete family of methods. Many different training methods are able to generate word embeddings from unstructured data, making the novel semantic analysis models achieve state-of-art performance. As for the legal point of view, it is critical to find out an efficient way to represent the semantic meaning of the longer pieces of text, such as phrases, sentences, or documents, to achieve a more reasonable interpretation. 

On the other hand, one major limiting factor in this context is that there are usually no legal datasets to train neural models at a scale properly. Hence, the performance expected for an LQA system is usually worse than in a generic one. 

\begin{tcolorbox}
\noindent \textbf{Pros:} These methods are capable of achieving the best results to date \\
\noindent \textbf{Cons:} Very poor interpretability. Huge amount of data needed for training
\end{tcolorbox}

As a final note, it is worth mentioning that although the lack of interpretability of neural solutions is always mentioned, an effort is being made in recent times to develop frameworks that help the human operator \citep{key-Lime,key-Shap}. These frameworks are intended to enable machine learning engineers and relevant domain specialists to analyze the end-to-end solutions and discover differences that could lead to a sub-optimal performance concerning the desired objectives. 

\subsection{Towards a new family of solutions}
Without detracting from the merit of neural solutions, many legal professionals are not satisfied with just an answer to their question and demand much more. They demand to understand why the solution has opted for such an answer and not another alternative. This brings up one of the ghosts that has always been associated with neural computing, that is, that its behavior is similar to that of a black-box since it is possible to give it an input and obtain an output. However, it is humanly impossible to understand how the connections of thousands of neurons have worked to give rise to such an answer, or even the real meaning behind the feature vectors obtained after subjecting the model to the training of a neural nature. For example, the base configuration of BERT \citep{key-BERT} requires a configuration consisting of 12 layers of neurons and 12 windows of attention. Therefore, there are usually issues related to the interpretability of the resulting model. This is where this family of methods comes in.

While Linked Data is now well-established and has a strong community behind it that has been studying it for several years, there is a lack of work on its application to the legal domain. It is true, however, that some proposals have begun to be put on the table. Nevertheless, what really promises to impact this area is the combination of Knowledge Graphs and Machine Learning to create a new generation of LQA systems.

To date, some authors thought that exploring the Linked Data route was the way to go \citep{key-He,key-Filtz}. However, the solutions are still unrealistic, as most of the legislation is still not published in a structured way. Furthermore, automatically structuring it is currently unfeasible. In addition, the learning curve for formulating questions in graph-oriented languages has led to the search for alternative approaches. In recent times, several works have also emerged that use for the first time the potential of knowledge graphs to overcome some of the gaps of neural solutions, such as the lack of interpretability and the need for large amounts of data for training \citep{key-Tong,key-Sovrano,key-Huang2,key-Filtz2}. In addition, technologies such as GPT-3\footnote{https://openai.com/} have been developed to convert questions formulated in natural language to their equivalent in graph-oriented language. Table 9 lists the works above.

\begin{table}[]
\centering
\caption{Summary of LQA approaches based on other approaches}
\begin{tabular}{ll}
\hline
\textbf{Work} & \textbf{Approach }\\
\hline \hline
\rowcolor{Gray}
\citep{key-He}     &     Linked Data      \\
\citep{key-Filtz}   & Linked Data  \\ 
\rowcolor{Gray}
\citep{key-Tong}   & Legal Knowledge Graph  \\ 
\citep{key-Sovrano}   & Legal Knowledge Graph   \\  
\rowcolor{Gray}
\citep{key-Huang2}   & Legal Knowledge Graph  \\ 
\citep{key-Filtz2}     &     Legal Knowledge Graph      \\
\hline     
\end{tabular}
\end{table}

Legal Knowledge Graphs has attracted much attention in recent times due to that an increasing number of researchers point out that they can bring a high degree of accuracy together with the highest interpretability that can be achieved to date. However, the other side of the coin says that this approach can offer good results at a high cost (in terms of money, time, and effort needed). This is mainly because this technology faces some obstacles in its development related to the amount of engineering work to build graphs properly and the necessary improvement of natural language translators to graph query language.

\begin{tcolorbox}
\noindent \textbf{Pros:} High accuracy. High interpretability. High performance \\
\noindent \textbf{Cons:} No good or cheap methods for automatizing the whole process
\end{tcolorbox}

\subsection{Summary}
It is clear from the literature that working with information concerning legislation and case law has always been attractive to computer scientists and practitioners applying for the latest advances in language and information technologies. These technologies have proven to be very useful for solving several problems that have traditionally affected legal information processing. In practice, the daily activities of these legal professionals requires checking vast amounts of legal material necessary to assess the relevant information pieces and identify the correct fragment they need. With this objective in mind, the LQA discipline was born and has been developed.

We provide a general summary of the different methods followed to build LQAs in the last decades. Figure 1 shows how, at present, of the 65 primary sources analyzed, the vast majority belong to the families of neural solutions and IR-based methods, although some families, such as other approaches, e.g., Legal Knowledge Graphs, are getting traction in recent times.  

\begin{figure}[h]
	\centering
	\begin{tikzpicture}	 
		\pie[
    color = {
        gray!80,
        gray!60,
        gray!70, 
        gray!50,
				gray!60, 
        gray!40, 
        gray!50}
		]{32.3/Neural Solutions,
    26.1/Information Retrieval,
    9.2/Yes-No Answers,
    9.2/LMCQA,
    9.2/Ontologies,
		9.2/Other approaches,
    4.8/Big Data}
 
	\end{tikzpicture}
\caption{Summary of existing families to meet the challenge of LQA}
\end{figure}
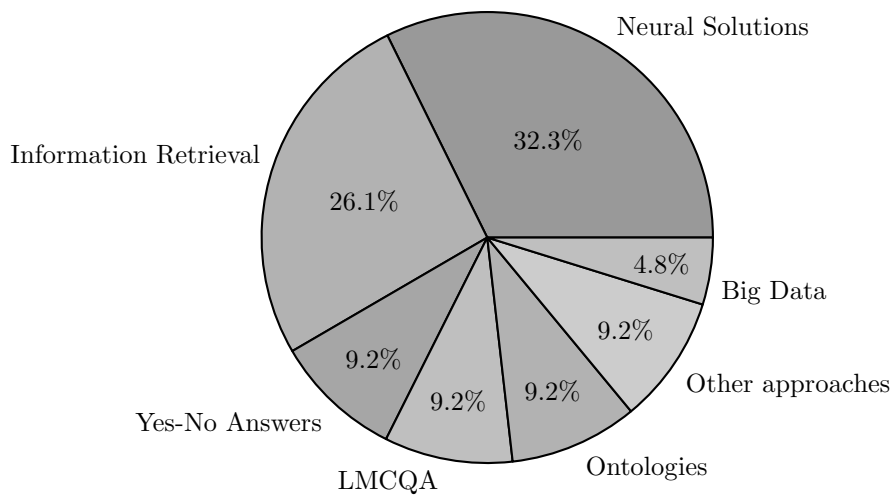

Next, we will look at the sources of data, information, and knowledge that are commonly used when building LQA solutions.

\section{Data, Information and Knowledge Sources}
LQA solutions are a kind of system that analyzes one or various data sources to answer a given question automatically. These data sources are usually structured or unstructured. The first ones are commonly referred to as legal corpora, while the latter are usually considered Legal Knowledge Bases (LKBs). Please note that in this domain, data is considered the atomic unit with which the systems work. If this data is structured to facilitate its interpretation, it is considered information. Finally, when relationships are created to facilitate understanding and communication, it is considered knowledge.

Depending on the sources to be exploited, two techniques for tackling the problem are working with unstructured legal corpora or working LKBs. Sometimes a third option is also considered concerning semi-structured sources (although this scenario is usually considered structured as well). Each one has its own set of benefits and drawbacks. Working with structured LKBs, for example, let system designers take advantage of the knowledge represented by using so-called inference engines to infer new information and answer questions. The fact is that not easy to implement these systems, so they have been progressively replaced by more efficient systems based on lighter knowledge models such as knowledge graphs and other enhanced lexical semantic models \citep{key-Yih}. However, at the beginning of this survey, we mentioned that legislation usually comes in an unstructured form, so it is more realistic to design QA systems being prepared to ingest vast amounts of data from unstructured sources. We will now focus on the two fundamental kinds of data sources that LQA solutions typically use to find the appropriate answers to the questions asked. We divide these sources into unstructured and structured.

\subsection{Unstructured sources}
As we have seen repeatedly in this survey, most legislation is generated in an unstructured manner and at high speed. This means that designing methods capable of working with unstructured text is usually the norm when dealing with legal information. LQAs solutions exploiting unstructured sources have several practical benefits as most of them have been specifically designed to efficiently process vast amounts of textual data (usually represented in legal language). These enormous amounts of data come from existing documents, legal databases, and so on. Most of the methods of the Information Retrieval and Neural Solutions families are based on this strategy.

The impossibility of having structured legal information on a large scale, of high quality, and covering all the current legislation, makes researchers and practitioners consider using methods to exploit unstructured resources of a legal nature. For this reason, the most frequent type of LQA solution needs to identify the relevant legal documents or information pieces and split them into candidate passages, and then, face the main key to the problem that consists of extracting the answer from the candidate passages.

The current generation of LQA solutions has evolved to extract answers from a wide range of different plain machine-readable resources. These LQA solutions exploit the massive set of unstructured information available on some sources to retrieve information about any particular question. It is important to note that these QA systems are possible mainly due to the extraction functions that are usually either heuristic or learned from a dataset such the Stanford Question Answering Dataset (SQuAD) \citep{key-Rajpurkar} or JEC-QA \citep{key-Zhong2} in this particular case. Moreover, since these extraction methods can process questions about different domains and topics, they are highly reusable.

Concrete examples in the legal domain are ResPubliQA 2010 collection, i.e., a subset of the JRC-ACQUIS Multilingual Parallel Corpus and the EUROPARL collection which are multilingual parallel collections. The JRC-ACQUIS is a legal corpus representing the complete body of the European Union documents, and it is commonly used to build systems. Other interesting examples are the House of Lords judgments corpus\footnote{https://publications.parliament.uk/pa/ld/ldjudgmt.htm}, etc. Moreover, some recent works such as \citep{key-Chalkidis} show us one of the clearest examples of how LQA solutions can benefit from unstructured corpora to build successful systems. 

\begin{tcolorbox}
\noindent \textbf{Pros:} Information can be easily added or updated \\
\noindent \textbf{Cons:} Reliability of answers is low, paraphrasing is frequent, problematic representation of unstructured data 
\end{tcolorbox}

\subsection{Structured sources}
The most popular form of the structured source is Legal Knowledge Bases (LKBs) \citep{key-Capon}. The LKBs are knowledge management systems being able to describe legal entities and the relationships between them. Legal entities are usually represented as nodes, while the relationships between them are represented as edges. Each legal entity can be classified using one or more classes. At the same time, classes can be organized hierarchically. The same is valid for relationships. Nodes also stand for literals. Moreover, most LKBs can perform reasoning enabling the derivation of inferred facts. Most of ontology-based, Linked Data, and Legal Knowledge Graphs use these kinds of structured sources. 

For example, an LQA solution consuming knowledge from a LKB might be focused on translating a question formulated by a user to a specific query of an LKB, such as in \citep{key-Kourtin}. This makes these LQA solutions perfect candidates for answering factoid formulations with high accuracy. The reason is that factoid formulations are simple questions that could be answered with a precise fact, so the corresponding answer can be obtained merely by knowing an entity and a property in the LKB. Nevertheless, many problems may arise when many facts need to be discovered, connected, and combined to elaborate.

Furthermore, there may be several reasons why the generation of structured information is not so popular in this domain, but in general, it is widely assumed that generating such information might be too expensive in terms of resource consumption, it might be subject to many errors what it would make it difficult and expensive to be maintained \citep{key-martinez-cosrev}.

Despite these problems, general-purpose datasets to test solutions have been developed in recent times. For example, the series of Question Answering over Linked Data (QALD) \citep{key-Cimiano}. However, and unfortunately, the development of structured sources of legal content is much more limited. Although the process on how the extraction function is usually learned is well described in \citep{key-Zou}. The good news is that the use of Knowledge Graphs of a legal nature is being investigated to obtain better results \citep{key-Filtz,key-Huang2,key-Filtz2}.

\begin{tcolorbox}
\noindent \textbf{Pros:} Reliability of answers is high, not need of complex natural language processing \\
\noindent \textbf{Cons:} Limited information stored, sourcing is error-prone and expensive to build, difficult interoperability in data sources 
\end{tcolorbox}

\section{The problem of the interpretability}
Historically, the legal sector is as interested in knowing the answer to a question as in knowing its associated explanation. It is something inherent to this domain that differentiates it from many disciplines where a black-box would have no significant problems operating. In the legal industry, apart from the scenario whereby the system is assumed to have the required information and the user is willing to accept the answers, another critical factor comes into play: the need for explanation. Therefore, the problem of interpretability is highly relevant for legal professionals.

Methods and techniques for answering specific questions are in high demand, and as a result, several solutions for LQA have been developed to respond to this need. System designers claim that the capability to answer questions through computers automatically could help alleviate a problem involving tedious tasks such as an extensive information search that is, in general, time-consuming. Moreover, therefore, by automatically providing hints concerning a vast number of legal topics, lots of resources in the form of effort, money, and time could be saved. 

The problem is that designing a LQA solution with a high degree of accuracy, interpretability, and performance at the same time is far from being trivial. So one often has to choose at most two characteristics out of the three possible ones. Interpretability being, in the legal sector, almost a mandatory requirement. In addition, if it is taken into account that, as we have seen, black-box solutions currently dominate this application domain, it is possible to deduce that one of the root causes that is limiting the LQA growth is the fact that the automatic generation of explanations is still not very satisfactory for the legal professionals.

There are many proposals to work with legal texts; some of these proposals are based on variations of the concept of the distributional assumption terms that appear in a similar context base, the calculation of synonyms, others are based on the co-occurrence of words in the same textual corpus, different neural architectures, different legal sources, etc. In principle, it is challenging to discern which approach could perform better than the others. This always depends on the use case and the context in which they are applied. But there is at least one factor that can be known in advance: the solution's interpretability. That is, a human operator can fully understand the model used to answer questions. A system that is worthy of people's trust must not only be effective and efficient; it must also be able to communicate the reason for the outputs it produces \citep{key-Atkinson}. 

If we did not have a fundamental requirement, the interpretability of the solution, it is clear that the family of neural solutions might be a perfect candidate to build the best LQA solutions to date. However, understanding the model, which is an essential aspect in the legal domain, is not the only obstacle. For example, finding large data sets representing solved cases is very difficult or expensive, or the neural model is usually difficult to be reused in problems of a similar nature. In this respect, some of the latest IR systems are superior because they do not require training, and they can always be reused over any corpora.

For all these reasons we have detailed, explaining the answers offered is very important. Moreover, explanations should be contrastive. This means that in addition to explaining why a particular outcome has been offered, a good explanation should communicate why other outcomes have been discarded. According to recent work \citep{key-Doshi}, the interpretability of a model is measured according to three levels:

\begin{itemize}
  \item \textit{Application level interpretability} is the level at which a domain expert could understand how the model works. Therefore, a good knowledge of the legal area and how the quality of the result can be determined is needed. This level should be sufficient for specialized professionals in the sector.

  \item \textit{Human level interpretability} is a more superficial level than the application level because it is not necessary to rely on an expert to examine the model but an average person. This makes the experiments cheaper and more straightforward because it is possible to find many more people for the validation phase of the model. In addition, it could help the widespread adoption of such solutions.
  
 \item \textit{Function level interpretability} does not need people. It is assumed that anyone could interpret the resulting model. For example, everyone can understand how decision trees work. However, the limitation of functional level interpretability when working with text is limited \citep{key-martinez-eswa}.
\end{itemize}

To start working, a system with good interpretability at the application level should be sufficient. At least for a professional in the sector. However, this is often not enough because not all professionals are specialized in the same legal field. So the aim here is to reach high levels of interpretability regarding the application level, human level, and function level at the same time, as well as the need for many fewer resources for training and the possibility of a simple transfer learning stage. If not, the user will not necessarily assume that the system has a good understanding of the domain, which could be used to explain the predictions. As a result, it will choose not to use it. 

This create big problems since the simplest regression-based models are easy for users to interpret. However, they do not lead to the best results. The opposite is true for models of a neural nature. Therefore, accuracy and interpretability are often considered orthogonal goals. Therefore, building a solution that reconciles the two features is far from being trivial.

To date, no fully interpretable LQA solution delivery highly accurate results have been achieved. However, the use of Legal Knowledge Graphs seems promising because they combine the computational power of machine learning with the structured nature of knowledge bases (in the form of graphs). Time will tell if this family of methods can deliver what it promises. In the meantime, users of highly accurate neural LQA solutions can count on some frameworks that allow them to get great clues about how systems make the decisions they do \citep{key-Lime,key-Shap}.

\section{Evaluation}
Several datasets \cite{key-Sovrano2} and performance metrics have been proposed to evaluate the results of a given QA system. According to Rodrigo et al. \citep{key-Rodrigo} the method to obtain the final score depends on the evaluation measure selected. Each measure evaluates a different set of features. Hence, researchers must select the measure depending on the objectives of the evaluation.

The most relevant metrics to evaluate the quality of LQA solutions are the following: Without	ranking, ranking by question, ranking for all questions, cardinal confidence self-score, and other metrics, for example, to evaluate the quality of the artificially generated text.  We are going to see them in detail below.

\subsection{Without	Ranking}
Although accuracy is a simple and intuitive measure, it only acknowledges correct answers and does not consider the amount of incorrect answers. However, its application is very popular because it is easy to understand.

\begin{equation}
	 accuracy = \frac{correct \ answers}{total \ answers}
\end{equation}

\subsection{Ranking by question}
When the system offers several answers in order of priority and what is of interest is to average the correct answer in the ordered list, the following Mean Average Precision (MAP) is usually used. The MAP compares the ground truth of correct answers to the answers offered by the system and returns a score. The higher the score, the more accurate the LQA solution is.

\begin{equation}
MAP = \frac{\sum_{q=1}^Q AveP(q)}{Q} 
\end{equation}

whereby $Q$ is the number of questions and $AveP$ is defined as 

\begin{equation}
AveP = \frac{\sum_{k=1}^n P(k) \times rel(k)}{\mbox{number of relevant documents}} 
\end{equation}

whereby $rel(k)$ is a function returning 1 if the answer is correct

It is also usual to use the Mean Reciprocal Rank (MRR). MRR is a statistic measure for evaluating any process that produces a list of possible answers to a set of queries, ordered by the probability of correctness.

\begin{equation}
	 MRR = \frac{1}{|Q|} \sum_{i=1}^{|Q|} \frac{1}{rank_i}
\end{equation}

When there is a list of questions, precision and recall are evaluated based on the complete list of known distinct instances of the answers. Precision is the fraction of answers that are correct in relation to the found ones:

\begin{equation}
	precision=\frac{\{found \ answers\}\cap\{correct \ answers\}}{\{found \ answers\}}
\end{equation}

The recall is the fraction of the found answers in relation to the correct ones.

\begin{equation}
	recall=\frac{\{found \ answers\}\cap\{correct \ answers\}}{\{correct \ answers\}}
\end{equation}

F-measure combines precision and recall into a harmonic mean to summarize both metrics.

\begin{equation}
	f-measure=\frac{2 \cdot precision \cdot recall}{precision + recall}
\end{equation}

Precision can be optimized at the expense of recall and vice versa. For this reason, it is convenient to report the two measures together, or at least, the combination of them through an F-measure.

\subsection{Ranking for all questions}
The most popular metric for this scenario is the so-called Confidence Weighted Score (CWS), which is based on the notion of average precision. CWS requires LQA solution to return only one answer per question and rank all the answers according to the background truth. Then, the CWS metrics reward solutions returning correct answers at top positions in the ranking.

\begin{equation}
CWS = \frac{1}{|Q|} \sum_{i=1}^{|Q|} \frac{C(i)}{i}
\end{equation}

whereby $Q$ is the number of questions and $C$ is defined as 

\begin{equation}
C = \sum_{j=1}^{i} I(j)
\end{equation}

being $I(j)$ the function that returns 1 if answer j is correct and 0 if it is not.

\subsection{Cardinal Confidence Self-Score}
The most popular metrics in this context are K and K1 \citep{key-clef2004}. K and K1 are based on a utility function that returns -1 if the answer is incorrect and 1 if it is correct. Both measures weigh this value with the confidence score given by the LQA solution. A positive value does not indicate more correct answers than incorrect ones, but that the sum of scores from correct answers is higher than the sum of scores from incorrect answers \citep{key-Rodrigo}.

\subsection{Unanswered different from incorrect}
In the context of LQA solutions, most of the time, it is preferable not to respond incorrectly \citep{key-Penas3}. This is because it is usually unsuitable for confusing the user with incorrect answers that undermine the system's trustworthiness. This idea is not new, but despite several previous attempts, there is no commonly accepted measure to assess non-response.

To deal with this situation, the metric Correctness at one c@1 has been proposed. 

\begin{equation}
c@1 = \frac{1}{total \ questions} \cdot (correct \ answers + \frac{correct \ answers}{total \ questions} \cdot unanswered \ questions)
\end{equation}

\subsection{Other metrics}
In recent times, new alternative metrics have emerged to evaluate the quality of responses \citep{key-Chen}. These metrics are of particular relevance for some families of LQA construction methods, especially those that aim to generate a textual sentence and not just answer a factual question. A summary of these alternative metrics is shown below.

\subsubsection{BERT Score and its variants} BERTScore is a family of alternative metrics that first tries to get the BERT embeddings \citep{key-BERT} of each term and calculate the current answer and the ground truth through a BERT model separately. Then, a mapping is computed between candidate and reference terms by means of pairwise cosine similarity. This mapping is then aggregated into precision and recall scores, and then into a harmonic mean.

\subsubsection{Sentence Mover's Similarity} SMS is another alternative metric based on mover's distance to evaluate textual information such as machine-generated text. Its application in LQA assessment is done by firstly computing an embedding for each sentence in an answer as an average of the ELMo embeddings \citep{key-elmo}. Then, a mapping function is used to obtain the distance of moving a candidate answer's sentences to match the reference answer.

\subsubsection{N-gram based metrics} These are metrics developed for evaluating machine translation, e.g. BLEU, ROUGE, etc. The key idea is to evaluate a candidate sentence by determining the n-grams in the candidate sentence that also appear in the reference sentence \citep{key-Chen}. In the context of LQA solutions, these metrics can be adapted to measure the quality of the generated answer in relation to the answer specified in the ground truth.

\subsection{Summary}
In the following, we show a summary of the different families of methods to build LQAs and the most appropriate ways to automatically determine their quality. Table 10 shows a tabular summary of these correspondences.

\begin{table}[]
\centering
\caption{Most useful evaluation methods to determine the accuracy of LQA solutions}
\begin{tabular}{ll}
\hline
\textbf{Family of Methods} & \textbf{Evaluation}\\
\hline \hline
\rowcolor{Gray}
Yes/No Answers    &    Acc, KK1, c@1     \\
LMCQA  &  Acc, PR, MAP, MRR, CWS, KK1, c@1     \\
\rowcolor{Gray}
Classical IR-based solutions  & Acc, PR, MAP, MRR, CWS, KK1, c@1, BScore, SMS, NGram\\
Big Data solutions &    Acc, PR, MAP, MRR, CWS, KK1, c@1 \\
\rowcolor{Gray}
Ontology-based solutions &  Acc, Recall, KK1, c@1  \\
Neural solutions	&   Acc, PR, MAP, MRR, CWS, KK1, c@1, BScore, SMS, NGram\\
\rowcolor{Gray}
Other approaches &  Acc, PR, MAP, MRR, CWS, KK1, c@1, BScore, SMS, NGram\\
\hline     
\end{tabular}
\end{table}

For example, metrics such as BERTScore, SMS, and N-Gram do not make sense in the Yes/No Answer or LMCQA family of methods since the answers are known in advance and are unnecessary to generate them. Alternatively, when working with ontologies, the measure of recall is much more important than precision. Because if the answer was wrong, it is due to the access to a wrong ontology model rather than to the reasoning algorithm used to obtain the answer. Moreover, ranking measures do not make much sense either.

\section{Open Research Challenges}
Given the state-of-the-art and the open problems that need to be investigated, it is possible to identify some significant gaps that should be filled to facilitate the adoption of LQA solutions at scale. Among the Open Research Challenges (ORC) that can be identified are the lack of proposals for transfer learning in the legal sector, the lack of solutions to deal with the complicated nuances of legal language, the improvement of the integration of distributed and heterogeneous data sources, as well as the importance of the multilingual capabilities of the different solutions. In the following, we review each of them.

\subsection{ORC1. Transfer Learning}
One important concept related to knowledge representation is Transfer Learning. These methods are based on the possibility of using a given model to export it to other scenarios of analogous nature. In other words, the knowledge representation that has been learned to complete one task can be generalized to help complete other tasks. A suitable knowledge representation method must determine which factors and features will be exploited and thus reuse them in another task. 

Most traditional machine learning approaches can be used to build models capable of addressing various problems given enough data and time. However, in practice, the amount of data and time available are frequently limited. For this reason, transfer learning has received considerable attention from such communities as Deep Learning and Big Data communities to date. These communities' problems are very resource-intensive in the form of data and time. This fact explains why strategies of this kind are often seen as a way to alleviate such issues.

To date, transfer learning methods in the field of LQA solutions have been only superficially explored. Only the work of Yan et al. has attempted to bring solutions to the table \citep{key-Yan}. It will be necessary for the community to intensify its research in this area in the near future. It is a great way to boost the accuracy of solutions while reusing training data and computational resources.

\subsection{ORC2. Nuances of legal language}
While in many application domains of QA systems, there is an apparent problem of ambiguity, i.e., most of the time, the questions are formulated so that a single statement may assume different meanings. In legal language, the problem is just the opposite. The language is so precise that it cannot give rise to multiple interpretations. Although this may seem a facilitator a priori, this fact brings a series of disadvantages that should be faced.

Classical stemming techniques cannot be freely used to obtain the root of the terms. Thus, for example, the sentence \textit{the contract is void} is very different from the sentence \textit{the contract is voidable}. Furthermore, even though the terms void and voidable contain the same root, a similar meaning could not be assumed.

This forces us to have optimal performance methods to really understand the meaning of the information being processed. In the near future, we will have to continue working in this direction. This would be facilitated by exploring legal texts' inherent characteristics to utilize these features for building LQA solutions. For example, properties such as references between documents or structured relationships in legal statements should be investigated since they are often a great help in processing those different nuances that legal language brings.

\subsection{ORC3. Distributed Knowledge}
We have seen that although approaches based on structured corpora often yield good results, it is often difficult to use them in practice mainly due to the cost when building such structures (i.e., it is expensive in terms of effort, money, and time needed) and it is often complicated to find experts with enough knowledge for curating them. Therefore, it is usually desirable to access all possible sources to maximize the chances of success in answering a question while minimizing the expenditure of resources.

The problem is that the different data representations in the sources make automatic aggregation, interlinking, and integration very difficult. Therefore, the development of efficient methods to properly access differently structured sources of legal nature, which can serve as a knowledge source for computer systems in search of the ideal answer to a given question, is a critical challenge that should be faced in the near future.

For these reasons, further research is necessary to make the task easier, by different interlinking corpora based on diverse matching methodologies or by transforming and properly integrating other data source types into novel knowledge models, e.g., linked data, knowledge, graphs, etc. using some semantic labeling techniques, or even creating a unique global representation such as the universal schema such as the one proposed in \citep{key-Riedel}.

\subsection{ORC4. Multilingualism}
Working with several languages simultaneously is a recurring problem, not only in the design of LQA solutions. Nevertheless, the truth is that in this domain, it is of fundamental importance since there is important legislation and regulations that are not always written in the same language spoken in the territories where they apply.

The good news in this regard is the latest advances in word vectorization, which brings with it the possibility of working with abstract representations that are language-independent \citep{key-Grave}. In this sense, representing words by vectors can be considered one of the main recent breakthroughs in deep learning for natural language processing \citep{key-Goldberg}. The benefits of adopting vectors are multiple since, in addition to working with abstract representations of words, they have other associated advantages, such as low computational complexity, because these vectors are efficient to compute using simple operations.

The advances in word representation have a substantial impact on semantic analysis. The future semantic analysis models of legal text will improve LQA performance thanks to various deep learning methods to build multilingual word embeddings from unstructured text. However, to achieve a better performance, it is necessary to explore practical approaches describing the meaning of textual pieces, such as sentences, paragraphs, or even documents, from a legal perspective.

\section{Conclusions}
QA technology is becoming an essential solution in many areas overloaded by the constant generation of large amounts of information. Automatically answering specific questions can contribute to alleviating the problem of dealing with those vast amounts of data. In the legal domain, good LQA solutions are in high demand mainly due to the problems of information overload that legal professionals have to face in their daily activities. As a result, several solutions for LQA have been developed as a response to that situation. The primary reason for that is that the capability to answer questions through computers automatically could help alleviate a problem involving tedious tasks such as an extensive information search that is, in general, time-consuming and error-prone.  

We have surveyed the different LQA solutions that have been proposed in the literature to date. To do that, we have proceeded with the analysis of the essential methods in LQA over the last decades. We have identified the data sources on which existing LQA solutions have been implemented to automatically answer questions of legal nature. We have seen that there is no such solution that can provide high accuracy, interpretability, and performance at the same time. Nevertheless, there are usually strategies capable of optimizing two of them at the expense of the other, being interpretability of significant importance in the legal domain since the community has always found more useful systems whereby a few hundredths more precision does not compensate for the other disadvantages of interpretability and performance. Furthermore, we have made an overview of the most popular ways to determine the quality of LQA solutions from a strictly quantitative point of view. Last but not least, we have made a review of the research challenges that are still open today and that will have to be faced by the community soon.

This concludes this survey of more than two decades of research efforts in the field of LQA. It is expected that soon, more and more proposals for highly accurate, interpretable, and efficient QA systems specialized in legal matters will see the light of day. It seems also reasonable to think that LQA solutions will not replace legal experts and their unique, specialized knowledge. However, all indications are that these solutions will undoubtedly and significantly transform the traditional delivery of legal services in the near future.

\section*{Competing interest}
The author has no competing interest to declare.

\section*{Acknowledgments}
This research work has been partially supported by the Austrian Ministry for Transport, Innovation and Technology, the Federal Ministry of Science, Research and Economy, and the Province of Upper Austria in the frame of the COMET center SCCH.

\bibliography{mybib}

\end{document}